\documentclass[11pt]{article}

\newfont{\tenmsb}{msbm10 scaled\magstep1}

\let\ssection=\section
\renewcommand{\section}{\setcounter{equation}{0}\ssection}
\usepackage{graphics,amssymb,amsmath,amsthm,amscd,array}
\usepackage{graphicx,subfigure}
\usepackage{makeidx}
\usepackage{bm}% bold math
\usepackage{dsfont}

\font\BBig=cmr10 scaled\magstep2

\font\small=cmr9

\def\parag{\hfil\break} %%%%% paragraph
\def\kikezd{\parag\underbar}
\def\IR{{\mathds{R}}}  %%%%% Reals
\def\smallover#1/#2{\hbox{$\textstyle\frac{#1}{#2}$}} %
\def\smallcirc{{\,\raise 0.5pt \hbox{$\scriptstyle\circ$}\,}}
\def\2{{\smallover1/2}}
\def\={{\!=\!}}

\def\const{{\rm const}}

\def\dAlembert{\vcenter {
    \hbox {\vrule height8pt width0.4pt depth0.0pt
           \vrule height8pt width7.2pt depth-7.6pt
           \vrule height8pt width0.4pt depth0.0pt
           \kern-8pt
           \vrule height0.4pt width8pt depth0.0pt
          \,}}}%--- Francisco's box          
          
\def\vD{{\vD}}

\def\and{\qquad\hbox{\small and}\qquad}

\def\vD{{}{rightarrow{D}}}

\def\parag{\hfil\break} %%%%% paragraph
\def\={\!=\!}
\def\const{{\rm const}}

\begin{document}
\setlength{\baselineskip}{15pt}

%%%%%%%%%%%%%%%%%%%%%%%%%%%%%%%%%%%%%%%%%%%
%%%%%%%%%%%%%%%% the text %%%%%%%%%%%%%%%%%
%%%%%%%%%%%%%%%%%%%%%%%%%%%%%%%%%%%%%%%%%%%
\title{
{\bf\BBig
\centerline{The non-linear Schr\"odinger equation}
\medskip
\centerline{and}
\medskip
\centerline{
the conformal properties of non-relativistic space-time}}
\bigskip
\bigskip
} %%%%% for the front page

%%%%%%%%%%%%%%%%%%%%%%%%%%%%%%%%%%%%%%%%%%%
%%%%%%%%%%%%% the author(s) %%%%%%%%%%%%%%%
%%%%%%%%%%%%%%%%%%%%%%%%%%%%%%%%%%%%%%%%%%%

\author{P.~A.~Horv\'athy
\footnote{email~:horvathy-at-lmpt.univ-tours.fr}
\quad and\quad J.-C.~Yera
 \\
Laboratoire de Math\'ematiques et de Physique
Th\'eorique 
\\
Universit\'e de Tours (France)
}

%%%%%%%%%%%%%%%%%%%%%%%%%%%%%%%%%%%%%%%%%%%
%%%%%%%%%%%%%%%% the text %%%%%%%%%%%%%%%%%
%%%%%%%%%%%%%%%%%%%%%%%%%%%%%%%%%%%%%%%%%%%

\maketitle

\begin{abstract}
{\it The cubic non-linear Schr\"odin\-ger equation 
where the coefficient of the nonlinear term 
is a function $F(t,x)$ only passes the 
Painlev\'e test of Weiss, Tabor, and Carnevale
only for $F=(a+bt)^{-1}$, where $a$ and $b$ are constants.
This is explained by transforming the time-dependent system into the 
constant-coefficient NLS by means of a time-dependent
non-linear transformation, related to the conformal
properties of non-relativistic space-time.
A similar argument explains the integrability of the NLS in
a uniform force field or in an oscillator background.}
\end{abstract}                                                                                                                                                                  

%%%%%%%%%%%%%%%%%%%%%%%%%%%%%%%
%%%%%%%%%%% the paper %%%%%%%%%

%%%%%%%%%%%%%%%%%%%%%%
%\section{Introduction}
%%%%%%%%%%%%%%%%%%%%%%

The recent upsurge of interest in non-relativistic
conformal symmetries \cite{Hen,LSZGal,moreconf,DHNC} directed
attention to their role in getting a deeper understanding, and in
physical applications \cite{Hen}. 
In this Note we add another example to the list. To be specific, we  explain some interesting properties of the 
non-linear Schr\"odinger equation  (NLS) using these
symmetries.

\goodbreak
%%%%%%%%%%%%%%%%%%%%%%%%%%%%%%%%%%%%%%%%%%%%%%%%%%%%%%%%%%%%%%%%
\section{The NLS with a position and time-dependent non-linearity}
%%%%%%%%%%%%%%%%%%%%%%%%%%%%%%%%%%%%%%%%%%%%%%%%%%%%%%%%%%%%%%%%

Le us study  the cubic NLS
\begin{equation}
iu_t+u_{xx}+F(t,x)\vert u\vert^2u=0,
\label{2.1}
\end{equation} 
where $u=u(t,x)$ is a complex function in $1+1$ space-time dimension.
Such an equation arises, for example, in some approaches to
the  Quantum Hall Effect  \cite{AGJPS}.

When  $F(t,x)$ is a
constant, this is the usual NLS, which is known 
to be integrable. But what happens, when 
 the coefficient 
$F(t,x)$ is a {\it function} rather then just a constant~?

A useful test of integrability is provided by
the \textit{Painlev\'e test  of Weiss, Tabor and Carnevale}
\cite{WTC}. (The procedure is reminiscent of the Frobenius' method used for ODEs).

Let us recall the definition and some properties. For a full account, the Reader is advised to consult \cite{W1}.
Consider a system
of partial differential equations (PDEs), and let us assume that its
solutions are given by a meromorphic function of the
complex veriables $z_1,\dots, z_n$.  The singularities
of such a function belong to a manifold (called the singular manifold) of dimensions
$2n-2$, given by equations of the form
$
\Phi(z_1,\dots,z_n)=0,
$ 
where the $\Phi$ are analytical.

Then our PDE  is said to have the 
\textit{Painlev\'e property} if
all of its solutions can be written, in a neighbourhood of
the singular manifold, as a generalized  Laurent series,
\begin{equation}
u(z_1,\dots,z_n)=\Phi^{\alpha}\sum_{j=0}^\infty 
u_j(z_1,\dots,z_n)\Phi^j,
\label{Lexpan}
\end{equation}
where $\alpha$ is a negative integer and the $u_j(z_1,\dots,z_n)$ s are analytical.
Then the Painlev\'e conjecture of WTC \cite{WTC} says that
\textit{a PDE which has the Painlev\'e property is integrable} i.e. can be solved by inverse scattering.

Inserting the expansion (\ref{Lexpan}) into our PDE fixes the value of $\alpha$, and then provides us with
recurrence relations for the functions $u_j$. For some value of $j$ called resonances, $u_j$ remains undetermined,
and the system has to satisfy consistency conditions. 
 
Truncating the series may provide us with a B\"acklund transformation \cite{W1}. For example, one can generate
Jackiw-Pi vortex solutions from the vacuum \cite{HYPRD}.

Returning to the NLS, below we show 

\kikezd{Theorem1}~:
{\it The generalized non-linear Schr\"odinger equation (\ref{2.1})
only passes the Painlev\'e test  of Weiss, Tabor and Carnevale}
\cite{WTC}
{\it if the coefficient of the non-linear term
is} 
\begin{equation}
F(t,x)=\frac{1}{a+bt}, 
\qquad
a, b=\const.
\label{2.2}
\end{equation} 
\kikezd{Proof}.
As it is usual in studying non-linear Schr\"odiger-type equations
\cite{W1,CLACO}, we consider Eqn. (\ref{2.1}) together with its complex conjugate ($v=u^*$),
\begin{equation}
\begin{array}{lll}
 iu_t+u_{xx}+Fu^2v&=&0,
\\[6pt]
-iv_t+v_{xx}+Fv^2u&=&0.
\end{array}
\label{2.3}
\end{equation}
This system will pass the Painlev\'e  test if $u$ et $v$ 
have generalised Laurent series expansions,
\begin{equation}
u=\sum_{n=0}^{+\infty}u_n\xi^{n-p},\qquad
v=\sum_{n=0}^{+\infty}v_n\xi^{n-q},
\label{2.4}
\end{equation}
($u_n\equiv u_n(x,t)$, $v_n\equiv v_n(x,t)$ and $\xi\equiv\xi(x,t)$)
in the neighbourhood of the singular manifold $\xi(x,t)=0$, 
$\xi_x\not=0$,
with a sufficient number of free
coefficients. Owing to a results of  Weiss, and of Tabor \cite{W1,TAB},
it is enough to consider
$ 
\xi=x+\psi(t).
$  Then $u_n$ and $v_n$ become functions de $t$ alone,
$u_{n}\equiv u_{n}(t),\
v_{n}\equiv v_{n}(t). 
$ 
Checking the dominant terms, 
$u\sim u_0\xi^{-p}$, $v\sim v_0\xi^{-q}$, 
using the above remark, we get
\begin{equation}
p=q=1,
\qquad
F\,u_0v_0=-2.
\label{2.6}
\end{equation}
Hence  $F$ can only depend on $t$.
Now inserting the developments (\ref{2.4}) of $u$ and $v$ into (\ref{2.3}),
the terms in $\xi^k$, $k\geq -3$ read
\begin{equation}
\begin{array}{ll}
& i\Big(u_{k+1,t}+(k+1)u_{k+2}\xi_t\Big)+(k+2)(k+1)u_{k+3}
 +F\Big(\sum_{i+j+l=k+3}u_iu_jv_l\Big)=0,
 \\[6pt]
& i\Big(v_{k+1,t}+(k+1)v_{k+2}\xi_t\Big)-(k+2)(k+1)v_{k+3}
 -F\Big(\sum_{i+j+l=k+3}v_iv_ju_l\Big)=0.
 \end{array}
 \label{2.7}
\end{equation}
(Condition  (\ref{2.6})  is recovered for $k=-3$).    
The coefficients $u_n$, $v_n$ of the series (\ref{2.3}) are
given by the system $S_n$ ($k=n-3$),
\begin{equation}
\begin{array}{ll}
&[(n-1)(n-2)-4]u_n+Fu_0^2v_n=A_n,
\\[6pt]
& Fv_0^2u_n+[(n-1)(n-2)-4]v_n=B_n,
\end{array}
\label{2.8}
\end{equation}
where $A_n$ et $B_n$ only contain those terms $u_i$, $v_j$ with
$i,j<n$. The determinant of the system is 
\begin{equation}
\det S_n=n(n-4)(n-3)(n+1).
\label{2.9}
\end{equation}
Then (\ref{2.3}) passes the Painlev\'e test if, for each $n=0,3,4$, 
one of the coefficients $u_n$, $v_n$ can be arbitrary. 
For $n=0$, (\ref{2.6}) implies 
that this is indeed true either
for  $u_0$ or $v_0$. 
For $n=1$ and $n=2$, the system (\ref{2.7})-(\ref{2.8}) is readily solved, yielding 
\begin{equation}\begin{array}{ll}
&u_1=-\frac{i}{2}u_0\xi_t,
\qquad
v_1=\frac{i}{2}v_0\xi_t,
\\[6pt]
&6v_0u_2=iv_{0,t}u_0+2iu_{0,t}v_0-\frac{1}{2}u_0v_0(\xi_t)^2,
\\[6pt]
&6u_0v_2=-iu_{0,t}v_0-2iv_{0,t}u_0-\frac{1}{2}u_0v_0(\xi_t)^2.
\end{array}
\label{2.10}
\end{equation}
$n=3$ has to be a resonance; using condition (\ref{2.6}),
the system (\ref{2.8}) becomes
$$
\begin{array}{l}
-2v_0u_3-2u_0v_3=A_3v_0,
\\[6pt]
-2v_0u_3-2u_0v_3=B_3u_0,
\end{array}
$$
which requires $A_3v_0=B_3u_0$. 
But using the expressions of $A_{3}$ and $B_3$, with the
help of ``Mathematica'' we find 
$$
2FA_3=u_0(F_t\xi_t-F\xi_{tt}),\qquad u_0F^2B_3=F\xi_{tt}-F_t\xi_t,
$$
so that the required condition indeed holds.

$n=4$ has also to be a resonance;  we find, as before,
$$
\begin{array}{ll}
2v_0u_4-2u_0v_4&=A_4v_0,
\cr
-2v_0u_4-2u_0v_4&=B_4u_0,
\end{array}
$$
enforcing the relation
$
v_0A_4=-u_0B_4.
$ 
Now using the expressions of
$v_0$, $u_1$, $v_1$, $u_2$, $v_2$ as functions of $u_0$, $F$,  
$u_3$, $v_3$,  ``Mathematica'' yields
$$
\begin{array}{ll}
6u_0F^2A_4=
&-F^2u_{0,t}^2-2iu_0^2F^2\xi_t\xi_{tt}+u_0F^2u_{0,tt}
+iu_0^2F\xi_t^2F_t-u_0Fu_{0,t}F_t
+2u_0F_t^2-u_0^2FF_{tt},
\\[8pt]
3u_0^3F^3B_4=
&-F^2u_{0,t}^2-2iu_0^2F^2\xi_t\xi_{tt}+u_0F^2u_{0,tt}
+iu_0^2F\xi_t^2F_t-u_0Fu_{0,t}F_t
-4u_0F_t^2+2u_0^2FF_{tt}.
\end{array}
$$
Then our constraint implies that 
\begin{equation}
2F_t^2-FF_{tt}=0.
\end{equation}
Thus
$\big(F^{-1}\big)_{tt}=0$, 
so that $F^{-1}(x,t)=a+bt$, as stated. 

For $b=0$,
$
F(t,x)
$
in Eqn. (\ref{2.1}) is a constant,
and we recover the constant-coefficient NLS with its known solutions.
For $b\neq0$, the equation becomes explicitly time-dependent.
Assuming,
for simplicity, that $a=0$ and $b=1$, it reads
\begin{equation}
iu_{t}+u_{xx}+\frac{1}{ t}\vert u\vert^2u=0.
\label{2.13}
\end{equation}

This equation can also be solved. Generalizing the usual
travelling soliton, let us seek, for example,
 a solution of the form
\begin{equation}
u_0(t,x)=%\exp 
e^{i(x^2/4t-1/t)}\,f(t,x),
\label{2.14}
\end{equation}
where $f(t,x)$ is some real function. Inserting the  Ansatz
(\ref{2.14}) into (\ref{2.13}), a routine calculation yields the soliton
\begin{equation}
u_0(t,x)=\frac{e^{i(x^2/4t-1/t)}\,}{\sqrt{t}}\,
\frac{\sqrt{2}}{\cosh\big[x/t+x_{0}\big]}.
\label{2.18}
\end{equation}
Interestingly, the steps leading to (\ref{2.18}) are essentially the same as
those met when constructing travelling solitons for the ordinary NLS ---
and this is not a pure coincidence~:

\kikezd{Theorem2}.
\begin{equation}
u(t,x)=\frac{1}{\sqrt{t}}\exp\Big[\frac{ix^2}{4t}\Big]\,
U\big(-{1/t},-{x/t}\big)
\label{2.19}
\end{equation}
{\it satisfies the time-dependent equation (\ref{2.13})
if and only if 
$U(t,x)$ solves Eqn. (\ref{2.1}) with $F=1$}.
\goodbreak
\vskip2mm
This can readily be proved by a direct calculation. Inserting (\ref{2.19}) 
into (\ref{2.13}), we find,
\begin{equation}
iu_{t}+u_{xx}+\frac{1}{ t}\vert u\vert^2u=
t^{-5/2}\exp\Big[\frac{ix^2}{4t}\Big]\,\bigg(
iU_{t}+U_{xx}+\vert U\vert^2U\bigg),
\end{equation}
proving our statement.

Our soliton (\ref{2.18}) constructed above comes in fact from the
well-known ``standing soliton'' solution of the NLS,
\begin{equation}
U_{0}(t,x)=\frac{\sqrt{2}\,e^{it}}{\cosh[x-x_{0}]}\ ,
\label{2.20}
\end{equation}
by the transformation (\ref{2.19}). More general solutions
could be obtained starting with the travelling soliton
\begin{equation}
U(t,x)=e^{i(vt-kx)}\frac{\sqrt{2}\,a}{\cosh[a(x+kt)]},
\qquad
a=\sqrt{k^2+v}.
\label{2.21}
\end{equation}

%%%%%%%%%%%%%%%%%%%%%%%%%%%%%%%%%%%%%%%%%%%%%%%%%%%%%%%%%%%%
\section{Non-relativistic conformal transformations}
%%%%%%%%%%%%%%%%%%%%%%%%%%%%%%%%%%%%%%%%%%%%%%%%%%%%%%%%%%%%

Where does the formula (\ref{2.19}) come from~?
To explain it, let us remember 
that the non-linear space-time transformation
\begin{equation}
D:\left(\begin{array}{c}
t\\ x
\end{array}\right)
\to
\left(\begin{array}{c}
-\displaystyle{1/t}\\
-\displaystyle{x/t}
\end{array}\right)
\label{3.1}
\end{equation}
has already been met in a rather different context, namely
in describing planetary motion when the gravitational ``constant''
changes inversely with time, as suggested by Dirac \cite{DIR}. Then
one shows that 
\begin{equation}
\vec{r}(t)=t\,\vec{r}^*\big(-{1/t})
\label{3.2}
\end{equation} 
describes  planetary 
motion with Newton's ``constant'' varying as 
$ 
G(t)={G_0}{ t},
$ 
whenever
$\vec{r}^*(t)$ describes ordinary planetary motion, i.e. the one
with a constant gravitational constant, $G(t)=G_0$ \cite{DGH}
\footnote{Curiously, the {\it same} transformation is used to transform  supernova explosion into implosion,
\cite{ORS,HH}.}.

The strange-looking transformation (\ref{3.1}) is indeed related to the
conformal structure of non-relativistic space-time \cite{DHNC,DGH,DBKP,Henkel}.
It has been noticed a long time ago \cite{JNH}, that the 
``conformal''
space-time transformations
\begin{equation}\left\{\begin{array}{lllllll}
\left(\begin{array}{c}t\\ x\end{array}\right)
&\to&
\left(\begin{array}{c}
T\\
X\end{array}\right)&=&
\left(\begin{array}{c} 
\displaystyle{\delta^2}t
\\
\displaystyle{\delta}\, x
\end{array}\right),
\qquad 
&0\neq\delta\in\IR 
&\hbox{dilatations}
\\[8pt]
\left(\begin{array}{c}t\\ x\end{array}\right)
&\to&
\left(\begin{array}{c}T\\ X\end{array}\right)
&=&
\left(\begin{array}{c}
\displaystyle\frac{t}{1-\kappa t}
\\[6pt]
\displaystyle\frac{x}{1-\kappa t}
\end{array}\right),
\qquad
&
\kappa\in\IR 
&\hbox{expansions} 
\\[8pt]
\left(\begin{array}{c}t\\ x\end{array}\right)
&\to&
\left(\begin{array}{c}T\\ X\end{array}\right)
&=&
\left(\begin{array}{c}t+\epsilon\\ x\end{array}
\right),
\qquad 
&\epsilon\in\IR 
&\hbox{time translations} 
\end{array}\right.
\label{3.4}
\end{equation}
implemented on wave functions according to
\begin{equation}
U(T,X)=\left\{
\begin{array}{l}
\displaystyle{\delta}^{1/2}u(t,x) 
\\[8pt]
(1-\kappa t)^{1/2}\exp
\Big[i\displaystyle\frac{\kappa x^2}{4(1-\kappa t)}\Big]u(t,x)
\\[8pt]
u(t,x)
\end{array}\right.
\label{3.5}
\end{equation}
permute the solutions of
 the free Schr\"odinger equation. 
In other words, they are {\it symmetries} for the free Schr\"odinger 
equation. (The generators in (\ref{3.4}) span in fact an ${\rm SL}(2,\IR)$ 
group; when added to the obvious galilean symmetry, 
the  Schr\"odinger group is obtained.
A Dirac monopole, an Aharonov-Bohm 
vector potential, and an inverse-square potential can also be included,
\cite{Jmon,DGH,DHP}).

The  transformation $D$ in Eqn. (\ref{3.1}) belongs to
this symmetry group: it is in fact 
 (i)  a time translation with 
$\epsilon=1$, (ii) followed by an expansion with 
$\kappa=1$, (iii) followed by a second time-translation with
$\epsilon=1$.
It is hence a symmetry for the free (linear) Schr\"odinger
equation. Its action on $\psi$, deduced from (\ref{3.5}), is precisely (\ref{2.19}).

The cubic NLS with non-linearity $F=\const$.
 is not more $SL(2,\IR)$ invariant \footnote{
Galilean symmetry can be used to produce further
solutions --- just like the travelling soliton (\ref{2.21}) can be 
obtained from the ``standing one'' in (\ref{2.20}) by a galilean boost.
Full Schr\"odinger invariance yielding expanded and dilated solutions 
can be restored by replacing the cubic non-linear term by
the fifth-order non-linearity $\vert\psi\vert^4\psi$.
These statements about non-invariance assume  restricting ourselves to certain representations, see \cite{StoHe}.}.
In particular, the 
 transformation $D$ in (\ref{3.1}), implemented as in Eq. 
 (\ref{2.19}) carries the cubic term into the
time-dependent term
$(1/t)\vert u\vert^2u$ --- just like Newton's gravitational potential
$G_{0}/r$ with $G_0=\const$. is carried into the time-dependent
Dirac expression $t^{-1}G_{0}/r$ \cite{DGH}.

Similar arguments explain the integrability of other NLS-type
equations. For example, electromagnetic waves in a non-uniform
medium propagate according to
\begin{equation}
iu_{t}+u_{xx}+\big(-2\alpha x+2\vert u\vert^2\big)u=0,
\label{3.6}
\end{equation}
which can again be solved by inverse scattering   \cite{CL}. % to yield
%\begin{equation}
%\exp\Big[2(\xi-\alpha t)x-4\big[\smallover1/3\alpha^2t^3-\alpha\xi 
%t^2+(\xi^2-\eta^2)t\big]+\theta_{0}\Big]
%\frac{2\eta}{\cosh\big[2\eta(x+2\alpha t^2-4\xi t-x_{0})\big]}.
%\end{equation}
 This is explained by observing that the potential term here
can be eliminated by  switching to a uniformly accelerated frame: 
\begin{equation}
\begin{array}{ll}
&u(t,x)=\exp\big[-i(2\alpha xt+\smallover4/3\alpha^2 t^3\big]U(T,X),
\\[6pt]
&T=t,
\qquad
X=x+2\alpha t^2.
\end{array}
\label{3.7}
\end{equation}
Then $u(t,x)$ solves (\ref{3.6}) whenever $U(T,X)$ solves the free equation
$%\begin{equation}
iU_{t}+U_{xx}+2\vert U\vert^2U=0.
%\label{freeNLS}
$%\end{equation}

The transformation (\ref{3.7}) is again related to the structure of
non-relativistic
space-time. It can be shown in fact [10] that the (linear) Schr\"odinger
equation
\begin{equation}
iu_{t}+u_{xx}-V(t,x)u=0
\label{3.8}
\end{equation}
can be brought into the free form
$
iU_{T}+U_{XX}=0
$
 by a space-time transformation $(t,x)\to(T,X)$
if and only if the potential is
\begin{equation}
V(t,x)=\alpha(t)x\pm\frac{\omega^2(t)}{4}x^2.
\label{3.10}
\end{equation}

For the uniform force field ($\omega=0$)
the required space-time transformation is precisely (\ref{3.7}).
For the oscillator potential ($\alpha=0$), one can use rather
Niederer's transformation~\cite{NJP,DHP}
\goodbreak
\begin{equation}
\begin{array}{lll}
u(t,x)&=&\frac{1}{\sqrt{\cos\omega t}}\,
\exp\big[-i\frac{\omega}{4} x^2\tan\omega t\big]\,
U(T,X),
\\[12pt]
&&T=\displaystyle\frac{\tan\omega t}{\omega}
\qquad
X=\displaystyle\frac{x}{\cos\omega t}.
\end{array}
\label{3.11}
\end{equation}
Then
\begin{equation}
iu_t+u_{xx}-\frac{\omega^2x^2}{4}u
=
(\cos\omega t)^{-5/2}\,\exp\big[-i\frac{\omega}{4}\tan\omega t\big]
\Big(iU_T+U_{XX}\Big).
\label{3.12}
\end{equation}
Restoring the nonlinear term allows us to infer
that
\begin{equation}
iu_t+u_{xx}+\Big(-\frac{\omega^2x^2}{4}+\frac{1}{\cos\omega t}\,
\vert u\vert^2\Big)u=0
\label{3.13}
\end{equation}
is integrable, and its  solutions are obtained from those of
the ``free'' NLS by the transformation (\ref{3.11}).

%%%%%%%%%%%%%%%%%%%%%
\section{Discussion}
%%%%%%%%%%%%%%%%%%%%%

To conclude, we us mention  
some more related results. 

Firstly, our result should be compared with the 
those of Chen et al. \cite{CLL}, who prove that the equation
\begin{equation}
iu_t+u_{xx}+F(\vert u\vert^2)u=0
\label{2.11}
\end{equation}
can be solved by inverse scattering
if and only if
$
F(\vert u\vert^2)=\lambda\vert u\vert^2,
$
 where $\lambda=\const$. Note, however, that Chen et al. only
study the case when the functional $F(\vert u\vert^2)$ 
is independent of the space-time coordinates $t$ and $x$.

It has also been shown that the non-linear Schr\"odinger equation
with time--dependent coefficients,
\begin{equation}
iu_t+p(t)u_{xx}+F(t)\vert u\vert^2u=0,
\label{4.2}
\end{equation}
can be transformed into the constant--coefficient form
whenever \cite{GR}
\begin{equation}
p(t)=F(t)\left(a+b\int^tp(s)ds\right).
\label{4.3}
\end{equation}
This same condition, which could also be obtaind by a suitable
generalization of our approach, was found later as the one needed for
the Painlev\'e test \cite{JO} applied to Eq. (\ref{4.3}). 

On  the other hand, the constant-coefficient, damped, driven NLS,
\begin{equation}
iu_t+u_{xx}+F(t)\vert u\vert^2u=a(t,x)u+b(t,x),
\label{4.4}
\end{equation}
was shown to pass the Painlev\'e test if
\begin{equation}
a(t,x)=\big(\2\partial_t\beta-\beta^2\big)+i\beta(t)
+\alpha_{1}(t)+\alpha_{0}(t),
\qquad
b(t,x)=0,
\label{4.5}
\end{equation}
\cite{CLA}, i.e., when the potential can be
transformed away by our ``non-relativistic conformal 
transformations''.

We only studied the case of 
$d=1$ space dimension. 
Similar results would hold for any $d\geq1$.
It is worth noting that more general dynamical symmetries of the NLS under subalgebras of the Schr\"odinger/conformal algebra were studied systematically 
   by S. Stoimenov and M. Henkel \cite{StoHe}.

At last, it is worth noting that the ``Kaluza-Klein-type''
framework, first proposed by Duval et al. \cite{DBKP,DGH}  has attracted
considerable recent attraction, namely in the
non-relativistic AdS/FCT context.  See, fore example, 
\cite{DHH}.

\vskip5mm\goodbreak
%%%%%%%%%%%%%%%%%%%%%%%%%%%%%%%%%%%%%%%%%%%
%%%%%%%%%%%%% the references %%%%%%%%%%%%%%
%%%%%%%%%%%%%%%%%%%%%%%%%%%%%%%%%%%%%%%%%%%

%%%%
\end{document}